\documentclass{aa}  

\usepackage{amssymb}
\usepackage{graphicx, natbib}
\bibpunct{(}{)}{;}{a}{}{,}
\usepackage[]{txfonts}
\begin{document}

   \title{Neutral interstellar He parameters in front of the heliosphere 1994--2007}

   \subtitle{}

   \author{M. Bzowski\inst{1}
          \and
          M.A. Kubiak\inst{1}
					\and 
					M. H{\l}ond\inst{1}
					\and
					J.M. Sok{\'o}{\l}\inst{1}
					\and
					M. Banaszkiewicz\inst{1}
					\and
					M. Witte\inst{2}
          }

   \institute{Space Research Centre of the Polish Academy of Sciences, Bartycka 18A, 00-716 Warsaw, Poland\\
              \email{bzowski@cbk.waw.pl}
         \and
             Max-Planck Institut f{\"u}r Sonnensystemforschung, Katlenburg-Lindau, Germany}

   \date{recived; accepted }

 
  \abstract
   {Recent analysis of IBEX measurements of neutral interstellar He flux brought the inflow velocity vector different from the results of earlier analysis of observations from the GAS instrument onboard Ulysses. Recapitulation of published results on the helium inflow direction from the past $\sim 40$ years suggested that the inflow direction may be changing with time.}   
   {We reanalyze the old Ulysses data and reprocess them to increase the accuracy of the instrument pointing to investigate if the GAS/Ulysses observations, carried out during almost two solar cycles, support the hypothesis that the interstellar helium inflow direction is changing.}
   {We employ a similar analysis method as used in the analysis of the IBEX data. We seek a parameter set that minimizes reduced chi-squared, using the Warsaw test particle model for the interstellar He flux at Ulysses with a state of the art model of neutral He ionization in the heliosphere, and precisely reproducing the observation conditions. We also propose a supplementary method of constraining the parameters based on cross-correlations of parameters obtained from analysis of carefully selected subsets of data. }
   {We find that the ecliptic longitude and speed of interstellar He are in a very good agreement with the values reported in the original GAS analysis. We find, however, that the temperature is markedly higher. The 3-seasons optimum parameter set is $\lambda = 255.3\degr$, $\beta = 6\degr$ (J2000), $v = 26.0$~km/s, $T = 7500$~K. We do not find evidence that these parameters are varing with time, but their uncertainty range is larger than originally reported.}
   {The originally-derived parameters of interstellar He from direct sampling on GAS/Ulysses are in good agreement with presently derived, except for the temperature, which seems to be appreciably higher, in good agreement with interstellar absorption line results. While the results of the present analysis are in marginal agreement with the earlier reported results from IBEX, the most likely values from the two analyses differ for reasons that are still not understood.}
   {}


   \maketitle
%

\section{Motivation}

The vector of inflow and temperature of neutral interstellar helium (NIS He) on the heliosphere were measured by two direct sampling experiments, GAS/Ulysses \citep{witte_etal:92a, witte_etal:93} and IBEX-Lo \citep{fuselier_etal:09b, mobius_etal:09a}, which returned different results. Based on measurements carried out from the middle of 1990-ties to 2001, GAS/Ulysses obtained J2000 $\lambda = 255.4\degr \pm 0.5\degr, \beta = 5.2\degr \pm 0.2\degr, v = 26.3 \pm 0.4$~km/s, $T = 6300 \pm 340$~K (\citet{witte:04}; see also \citet{witte_etal:96, witte_etal:04a}), while IBEX found $\lambda = 259.2\degr$ or 259.0\degr, with an uncertainty of $\sim 3\degr$, and $\beta = 5.1\degr$ or $4.9\degr, v = 22.8$ or 23.5~km/s, $T = 6200$ or $5300 - 9000$~K from measurements carried out in 2009 and 2010 (\citet{bzowski_etal:12a} and \citet{mobius_etal:12a}, respectively). The parameter values obtained from IBEX are tightly correlated with the inflow longitude and acceptable parameter ranges formed long alleys in the parameter space. A follow-up IBEX analysis brought consensus values suggested by \citet{mccomas_etal:12b}: $\lambda = 259.0\degr, \beta = 5.0\degr$ (J2000), $v = 23.2$~km/s, $T = 6300$~K. Based on these and other measurements from the past $\sim 40$~years, \citet{frisch_etal:13a} suggested that the flow vector of interstellar gas in the solar neighborhood might be changing. The existence of such a trend would have profound consequences for understanding the physics of the heliosphere and the  processes operating in the interstellar matter surrounding the Sun. 
	
Since the results from Ulysses and IBEX have far-reaching and different consequences for the physics of the heliosphere even without the variation of gas parameters in time\citep{mccomas_etal:12b, zank_etal:13a}, it is important to understand the reason for the differences, in particular, if the change in time, previously not considered in the analysis, can be ruled out. To that end, and in-depth re-analysis of the GAS/Ulysses data using a similar methodology and numerical models would be welcome, including the data from the last season of GAS observations, which up to now have not been analyzed in-depth. 

In the following, we analyze all seasons of GAS/Ulysses data. In addition to the data analyzed originally by \citet{witte:04}, we process the data from the last Ulysses season in 2007, less than two years before the IBEX launch. The analysis is carried out in a similar way to the analysis shown by \citet{bzowski_etal:12a}, using the same numerical code, adapted for the needs of GAS analysis, and using the same model of He ionization \citep{bzowski_etal:13b}. Other assumptions are also identical, including the important assumption that the whole signal can be approximated by a single, homogeneously distributed population of interstellar gas given by the Maxwell-Boltzmann distribution in front of the heliosphere. 


\section{GAS/Ulysses observations}

   \begin{figure*}
   \centering
   \resizebox{\hsize}{!}{\includegraphics[angle=90]{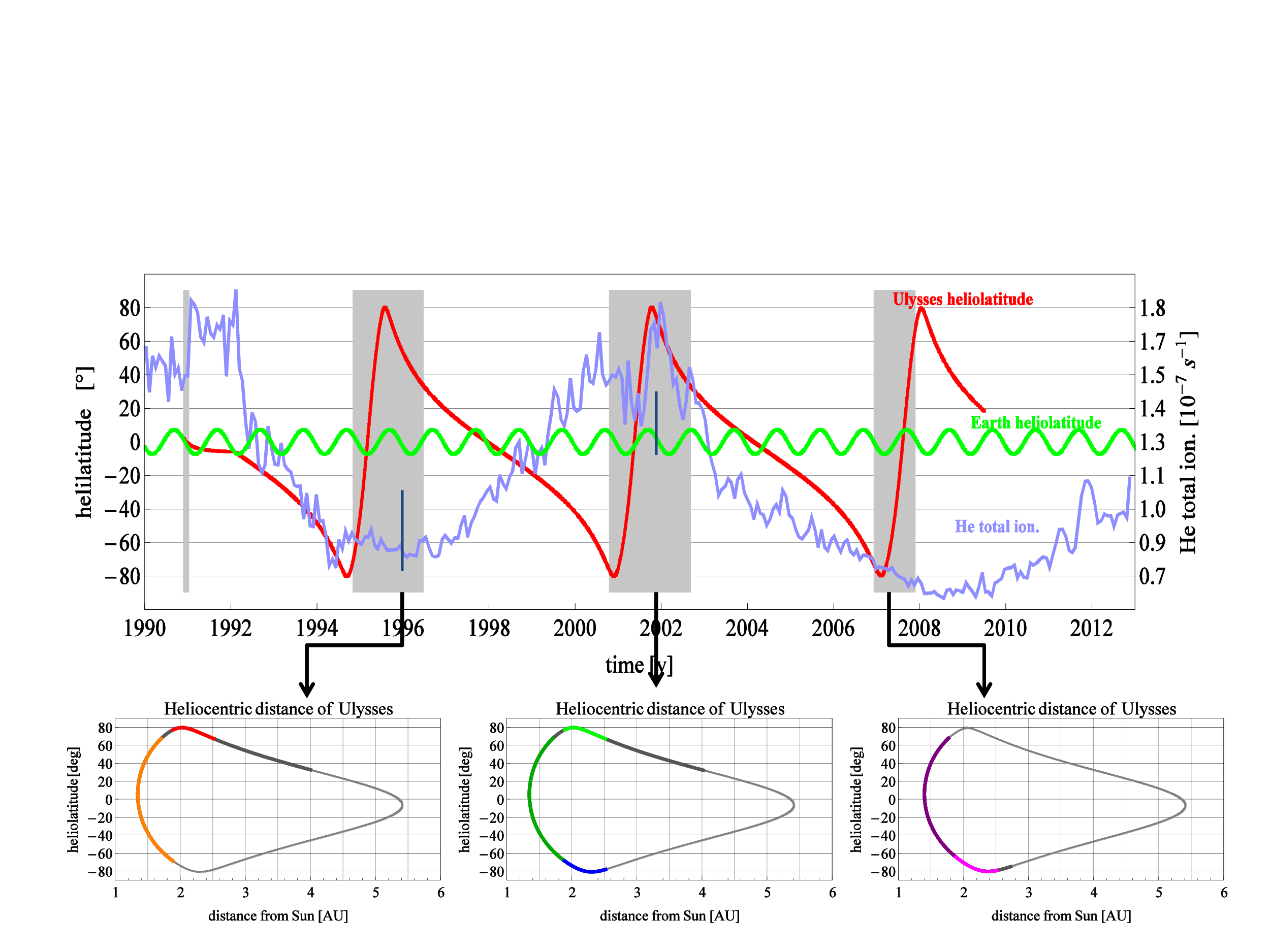}}
   \caption{Resume of GAS/Ulysses observations. The upper panel presents the Ulysses heliolatitude since launch until the end of mission in 2009 (red line, left-hand vertical scale). The green line shows the Earth heliolatitude. The blue line (right-hand scale) illustrates the helium photoionization rate used in this study, adopted after \citet{bzowski_etal:13b}. The dark-blue vertical bars mark the ionization rates (with uncertainties) obtained by \citet{witte:04} in his original analysis of the GAS observations. The gray regions mark the three GAS observation seasons on the polar orbit and the observation time during the in-ecliptic cruise phase shortly after launch. The three lower panels show the heliolatitude – solar distance hodograph of Ulysses during the three polar orbits. The dark gray regions mark the whole seasons taken to the analysis, and the colored arcs superimposed on the dark-gray line mark the arcs discussed farther in the text. The sum of the colored arcs is the entire data set for a given orbit. The orange, dark green, and dark-purple arcs mark the common arcs between the three  polar orbits. In the further part of the paper they are labeled CP95, CP01, and CP07, respectively. The red and light-green arcs mark the north polar scans, performed on the first and second polar orbit (NP95 and NP01, respectively), and the blue and magenta arcs mark the two south polar passes (SP01, SP07), executed on the second and third orbit. The boundaries of these arcs were carefully selected so that the heliolatitude ranges match between the orbits.}
              \label{figUlyCombo}%
    \end{figure*}
		
Ulysses was the first spacecraft to orbit the Sun in a polar plane. Launched on October 6, 1990, it first cruised to Jupiter to use its gravity assist in February 1992 to leave southward the ecliptic plane, and subsequently carried out almost three revolutions around the Sun on a nearly polar, elliptical orbit with the apogee at $\sim 5.5$~AU and perigee at $\sim 1.3$~AU from the Sun. The plane of the orbit was close to perpendicular to the inflow direction of interstellar gas. The trajectory of the spacecraft as a function of heliolatitude and distance from the Sun is shown in Fig.~\ref{figUlyCombo}. 

Ulysses was a spin-stabilized spacecraft, rotating at $\sim 5$~rpm, oriented so that its antenna always pointed towards the Earth. Thus the spin axis orientation was changing throughout the flight. The GAS instrument \citep{witte_etal:92a} provides two ``pin-hole'' cameras, which differ only in the width of their fields-of -view (FOV) for low and high angular resolution. The transmission functions of the collimators are given in Fig. 6 in \citet{witte_etal:92a}. The cameras are mounted on an instrument-provided stepping platform that allows to rotate their viewing direction to various ``elevation'' angles (or cone angles) with respect to the spin axis. The applicable coordinate system is shown in Fig.~11 in \citet{witte_etal:92a}. Harnessing also the spinning of the spacecraft, it was possible to scan strip-wise the whole sky or smaller selected areas of interest (e.g., around the direction of the inflowing interstellar gas atoms). The region of the sky to be observed was set in-flight by telecommand. The commanding was carried out in the spacecraft (rotating) reference system, which was tied to the celestial system by spin pulse time and attitude information, generated by the orientation system of the Ulysses spacecraft. In our analysis we used only the low-resolution data, which have higher signal-to-noise ratios. 

Having gone through the collimators, the atoms were hitting lead-glass surfaces covered with lithium fluoride (LiF) and sputtering Li$^+$ ions, which were registered by channeltron detectors. The LiF layer was periodically replenished in-flight in attempt to maintain the sensitivity of the instrument on a more or less constant level. In addition to the ion mode (the so-called NG mode), the instrument could be operated in the electron mode (the so-called UV mode), in which electrons sputtered by the heliospheric EUV radiation were registered. In addition to the sensitivity to He atoms, the instrument featured some remnant sensitivity to photons also in the ion mode. Thus, in addition to the charged particle background, it registered also some light from the heliospheric EUV backscatter glow and from EUV-bright stars. The latter ones were used a posteriori to better adjust the instrument pointing on the sky. 

The energy sensitivity of the instrument was a strong function of energy (see Fig.~8 in \citet{witte_etal:99a} and Fig.~1 in \citet{banaszkiewicz_etal:96a}). The process of calibration was complex and in the low energy range the calibration was known only to $\sim 20$\%, with two possible calibration functions. In our work we have approximated both calibration dependences with approximate analytic functions, with fit residuals better that $\sim 20$\% for the energies from 30 to 70~eV, relevant for the GAS experiment. Anticipating the results, careful analysis showed that owing to our data fitting approach, the differences between possible calibration functions turned out not to be a factor in our results.

Because of the energy limit in the sensitivity, the measurements could only be carried out on the perihelion portion of the orbit, where the relative speed of the spacecraft and incoming He atoms was the largest (see Fig.~\ref{figUlyCombo}). The instrument scanned selected areas on the sky, producing complete images of the distribution of the beam of interstellar He gas. Typically, there was one frame per one or two days, but some of them were unusable because of excessive background due to penetrating energetic particles from solar flares or other effects.

\section{Analysis}
In this paper we have used both the original data set and a new data set, obtained from reprocessing the old one. The processing of the original data set was described by \citet{witte_etal:93, witte_etal:96, banaszkiewicz_etal:96a, witte_etal:04a} and \citet{witte:04} and will not be repeated here. In the following, we present reprocessing the original data for this study, and subsequently detail the preparation of both the original and reprocessed data set for the parameter search we performed. 

\subsection{Reprocessing the original data}

\begin{figure*}
   \centering
   \includegraphics{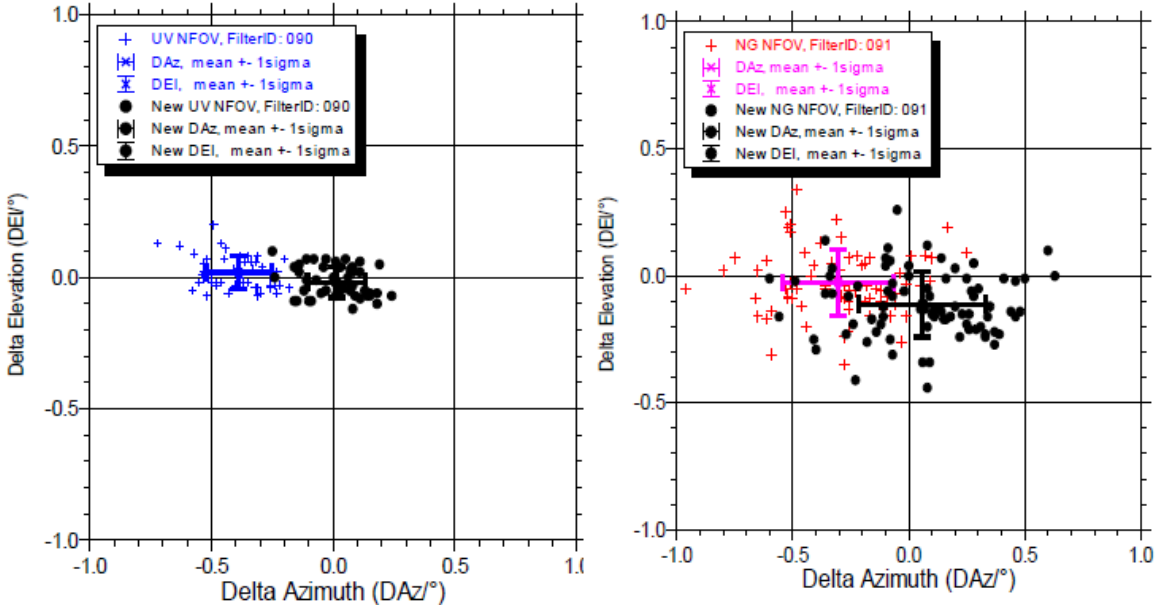}
   \caption{Scatter of the offsets that had to be applied to the matrices of transformation from the spacecraft to celestial coordinates, for observations made in the UV more (left panel) and in the NG mode (right panel) before (colored symbols) and after (black symbols) the adjustment of the original orientation calibration by $-0.4\degr$ in azimuth and $+0.06\degr$ in elevation. The crosses mark the one-sigma ranges of the respective distributions.}
    \label{NewAll_UV_N_090_091}
\end{figure*}
The main purpose of this reprocessing was to eliminate imperfections of the absolute pointing of the instrument on the sky and removing contamination of the signal from known stars and of excessive background frames in the subset from the last polar orbit of Ulysses, previously not analyzed. 

In principle, the only observable that can be measured directly by the GAS detector is the local arrival direction of interstellar neutral atoms at various positions along the spacecraft orbit. The physical parameters of NIS He can then be derived using the fact that the trajectory of the particles is given by a hyperbolic orbit in a plane containing the Sun as gravitational center and the asymptotic velocity vector at infinity \citep{witte_etal:92a}. This approach, simple for a cold (temperature = 0 K) NIS flow, becomes more complicated for a warm gas ($T >> 0$) and requires a sophisticated modeling using the classical hot model approach \citep[e.g.,][]{fahr:78}, as described by \citet{banaszkiewicz_etal:96a}, or an even more complex one, as used in our paper.

All this emphasizes the importance of a precise calibration of the viewing direction of the telescopes in the celestial (i.e., ecliptical) coordinates. The first end-to-end calibration was performed by \citet{witte_etal:04a} using UV-star data obtained during the 1995 season, mainly in NG-mode. In an end-to-end test, the pixel coordinates (in azimuth and elevation) of a star image were transformed into celestial coordinates and the resulting star position was compared with the well-known astronomical position obtained from a star catalog. Subsequently, a potential offset was used to adjust the transformation function until the observed vs catalog position offset was minimized. Small corrections of 0.55\degr in azimuth and $-0.16\degr$ in elevation were applied in this first calibration. It turned out that the values of the physical parameters, in particular the velocity value, significantly changed when the pointing calibration was adjusted by such small values, stressing the importance of an accuracy of that order \citep{witte_etal:04a}.

Now, in order to rule out calibration errors as a reason for the discrepancy between the GAS and the IBEX results, a further in-depth calibration was performed by the PI of the GAS experiment. This recalibration is now based on a much larger data basis, including the UV-star observations obtained during the seasons 2001 and 2007. In total there were 124 star images available, 75 taken in the NG-mode (with up to 400 counts per pixel) and 49 in the UV-mode (with up to 150\,000 counts per pixel). This dataset was subdivided into several groups in order to look for any systematic dependence of the offsets from one or more of the following parameters used during the original data taking:

{\noindent \em instrumental parameters (modes of operation)}
\begin{itemize}
\item[a)]{UV mode or NG mode}
\item[b)] {angular resolution of the data frame: pixel size in elevation (1\degr, 2\degr, or 4\degr) and in azimuth (0.7\degr, 1.4\degr, 2.8\degr)}
\end{itemize}
{\noindent \em spacecraft related (the stability of sun pulse, of the spin-sector pulses)}
\begin{itemize}
\item[a)]{solar aspect angle (the angle between the spin axis direction and the Sun)}
\item[b)]{ConScan operation (autonomous thruster firings executed by the orientation system of the spacecraft to stabilize the spin axis direction)}
\end{itemize}

{\noindent \em orbit-related}
\begin{itemize}
\item[a)]{south pole}
\item[b)]{center part}
\item[c)]{south pole}
\end{itemize}

Conclusive and statistically significant offset variations were seen only in the azimuth/elevation coordinates. An example is shown in Fig.~\ref{NewAll_UV_N_090_091}. The scatter in the NG-mode is larger than in the UV mode, mainly because of the much poorer statistics in the NG mode. The original pointing calibration was mainly based on the observations in the NG mode, which were frequently available as a byproduct of neutral gas observations. Presently, we can improve on the calibration on the basis of measurements in the UV mode and adopt the above correction values also for the measurements of neutral interstellar gas. 

As a result of this insight, for the reprocessed data set all NIS He and star data have been reprocessed with the adjusted pointing calibration and used in the present analysis in parallel with the original data set. 

\subsection{Data preparation}
The preparation of the data for parameter searching was carried out identically for the original and reprocessed data sets. The data are organized in data frames corresponding to individual “snapshots” (frames) of the neutral helium beam. Basically, at each location in the heliosphere NIS He gas features two beams, the so called direct and indirect beams. Along the Ulysses orbit, the indirect beam includes the atoms that have passed the perihelion of their orbits and thus the total intensity of this beam is greatly reduced because of ionization losses, which makes it hard to observe against background. Therefore in our analysis we used only the observations of the direct beam. 

In the first step, the background was determined on a frame by frame basis. With all data points exceeding 100 counts per 100 seconds rejected as certainly either being the signal or contaminated by EUV-bright stars, the background was calculated as the frame median. After testing a few different methods, it was ultimately concluded that the median and mode (the most probable value) from the histogram of all measured pixels in a given frame returned very similar values. The uncertainty was adopted as resulting from the Poisson distribution of the number of counts in a given pixel. Subsequently, data points were qualified for analysis using the criterion that they exceed the adopted background level by three standard deviations of the background level. In the next step, the background was subtracted, with the absolute uncertainty of the remaining data points unchanged. The few remaining “hot pixels”, disjoint from the signal peak and most likely being contaminated by EUV stars, were manually eliminated. This procedure was repeated separately for each data frame. 

The original data were specified in the ecliptic B1950 coordinates. They were converted to J2000 epoch and re-cast to spacecraft coordinates, since the parameter fitting was carried out in the Ulysses-inertial reference system. In the last step, the data and their uncertainties were re-normalized by dividing each pixel by the sum of all pixels kept in a given data frame. 

\subsection{Simulations}
\begin{figure}
{\includegraphics[width=10.0cm]{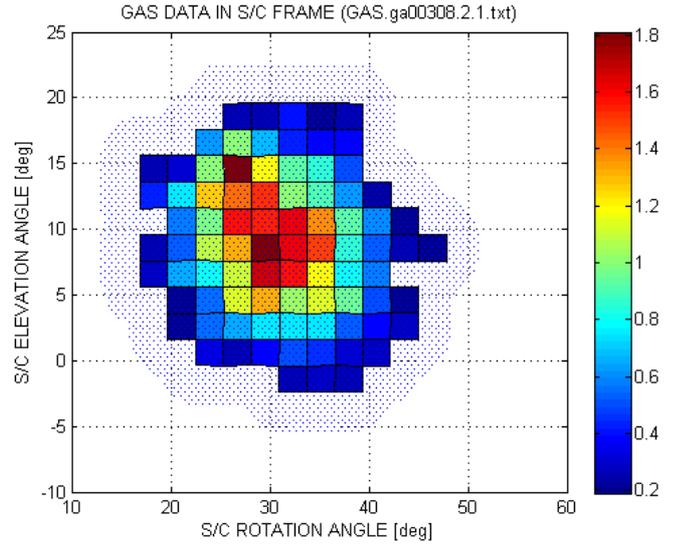}}
\caption{Example simulation grid for one of the data frames (tiny blue dots), superimposed on the data frame. In the simulation, first the NIS He flux grid was calculated in the Ulysses inertial frame, then the data pixels were simulated by numerical synthesizing the pixels by sliding the virtual collimator across the data frame similarly as it was done during the actual observations. The vertical scale is in counts per second.}
\label{figMesh}
\end{figure}

The simulations were carried out using specially adapted version of the Warsaw Test Particle Model \citep{tarnopolski_bzowski:09}, used by \citet{bzowski_etal:12a} for analysis of IBEX observations reported in \citet{mobius_etal:09b, bzowski_etal:12a}. The model is a kinetic test-particle code, equivalent to the hot model of interstellar gas, with finite temperature and time- and latitude-dependent ionization rate (for details of the hot model of neutral interstellar gas in the heliosphere see \citet{thomas:78} or \citet{fahr:78}, for details of the effects of time dependence in the ionization rate on the distribution of interstellar He in the heliosphere cf \citet{rucinski_etal:03}). In the simultions, we used  a time-and heliolatitude-dependent model of photoionization and electron-impact ionization rates based on actual observations of the solar EUV radiation and solar wind \citep{bzowski_etal:13b}. The actually used photoionization rate for the solar equator is shown as blue line in Fig.~\ref{figUlyCombo}. Out of ecliptic, we used an approximate formula used by \citet{bzowski_etal:12b} (see their Eq.~5) based on analysis by \citet{auchere_etal:05c}. The photoionization rate is quasi-elliptic as a function of heliolatitude, with polar rates less by approximately 20\% than the equatorial rates. In addition to photoionization, we took into account charge exchange with solar wind protons and alpha particles (negligible most of the time) and electron-impact ionization according to \citet{bzowski_etal:13b}, which vary with heliolatitude proportionally to solar wind density \citet{sokol_etal:12a}. 

Similarly to the analysis by \citet{bzowski_etal:12a}, the simulations were carried out in the spacecraft inertial frame. The first step was calculating the incident NIS He flux on a dense mesh (see Fig.~\ref{figMesh}) specially tailored for individual data frames, separately for all data frames. Next, a virtual collimator scanned the mesh, approximately reproducing the motion of the actual collimator on GAS so that the pixels with the centers corresponding to the data pixels were obtained. In the final step, renormalization was carried out similarly as for the data, i.e., the signal was divided by the sum of the calculated signals from all pixels.
The NIS He gas inflow parameters (the velocity vector and the temperature, together 4 independent parameters) were fitted by minimization of chi-squared for various assumed longitudes of the inflow direction, identically as did \citet{bzowski_etal:12a} (see their Eq.~16). For each of the assumed longitude values, the chi-squared minimum was searched for by varying the remaining parameters, i.e., the latitude of the inflow ecliptic longitude and latitude, speed, and temperature. 

We point out that because of the relatively large uncertainty in the absolute calibration on one hand, and expected large variations in the relative energy of the NIS He flux in the spacecraft frame on the other hand, we eliminated fitting of the absolute level of the signal from the optimization scheme. In our scheme, we fit the shape of the signal for all frames and require that all data frames in a given set taken for analysis match the fitted model. Potentially, such a scheme might end up in a seemingly satisfying solution that would be massively off in the terms of reproducing the total flux level. In further part of the paper we show that our approach has not produced such unwanted effects. 

\section{Results}
The goal of the study was on one hand verifying the original results by \citet{witte:04} using a different analysis method, and on the other hand checking if any traces of temporal change in the inflow parameters can be detected. To that end, we analyzed the data either in their entirety or split into various subsets. We made the fits both to the original data and to the data reprocessed for the needs of this analysis. 

\begin{figure}
{\includegraphics[width=9.0cm]{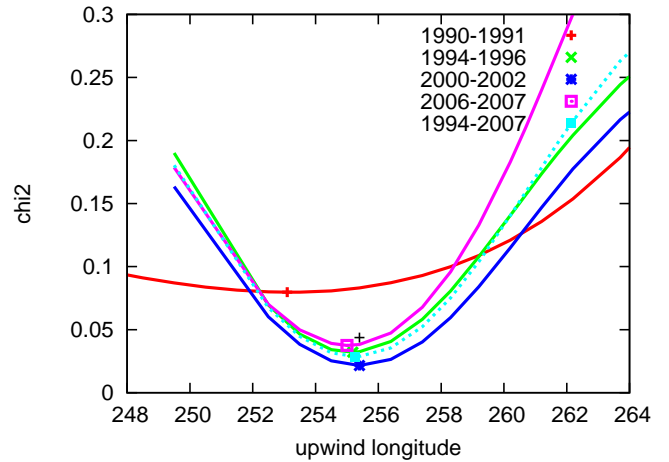}}
\caption{Chi-squared as a function of assumed inflow longitude (J2000) for the Ulysses cruise phase, three polar orbits fitted separately (solid lines), and the three polar orbits fitted together (broken line). The optimum solutions correspond to the minima. The chi-squared value obtained from a simulation of all three polar orbits with the parameters taken from \citet{witte:04} is marked with the black cross sign. The poorly-matching red line is obtained from the very few data frames collected shortly after launch, during the in-ecliptic cruise phase to Jupiter.}
\label{chi.longg}
\end{figure}

The NIS He inflow parameters obtained for the cruise phase in 1990-1991 and for each of the three full polar-orbit observation seasons are collected in Table~\ref{tabMinChi2Oldseasons} and chi-squared for these seasons as a function of inflow longitude is shown in Fig.~\ref{chi.longg}. Chi-squared minimization for the data from all three Ulysses polar orbits together (the global fit) yields $\lambda = 255.3\degr$, $\beta = 6.0\degr$, $v = 26.0$~km/s, $T = 7467$~K. These results are directly comparable with the parameters obtained by \citet{witte:04} (after converting the latter ones to J2000) since they were obtained using exactly the same data, without reprocessing; note, however, that \citet{witte:04} did not interpret the data from the last orbit.

\begin{table}
\caption[]{Interstellar He inflow parameters obtained from chi-squared minimization carried out separately for the ecliptic cruise and three polar orbits.}
\label{tabMinChi2Oldseasons}
\centering
\begin{tabular}{l c c c c}	
\hline \hline
    & $\lambda$[\degr] & $\beta$[\degr] & $v$~[km/s] & $T$[K] \\
\hline
cruise   & 252.2 & 4.23 & 27.100 & 10\,031 \\
1994--96 & 252.2 & 6.10 & 26.092 &  7538   \\
2000--02 & 255.4 & 5.80 & 26.087 &  7545  \\
2006--07 & 255.0 & 6.20 & 25.696 &  7209  \\
1994--07 & 255.3 & 6.00 & 25.970 &  7467  \\
\hline
\end{tabular}
\tablefoot{The last row shows the parameters obtained from fitting all data together}
\end{table} 

It is evident that the results of the fitting we have obtained now agree well with the original determination by \citet{witte:04} except for the temperature, which we obtained more than 1000~K higher: $6300 \pm 340$~K from \citet{witte:04} vs our $\sim 7500$~K. There are also other subtle differences: our inflow longitude is 255.3\degr, while the longitude from \citet{witte:04} converted to J2000 was $255.4\degr \pm 0.5\degr$. Our inflow latitude is a little higher: 6.0\degr vs $5.2\degr \pm 0.2\degr$. The speeds agree well: 26.0 vs $26.3 \pm 0.4$~km/s. 

The only significant departure from the parameters obtained from the global fit was the result for the in-ecliptic cruise phase, as shown in Table~\ref{tabMinChi2Oldseasons}. But the number of data frames collected during this phase was low and some spacecraft orientation issues on this very early phase of the mission were not fully resolved, so we do not consider this result as a severe inconsistency with the results obtained for the polar orbits. 
These results differ significantly from the results obtained from IBEX observations by \citet{bzowski_etal:12a} and \citet{mobius_etal:12a}, even after reconciliation of some small differences between these authors by \citet{mccomas_etal:12b}. Both \citet{bzowski_etal:12a} and \citet{mobius_etal:12a} pointed out, however, that acceptable parameters obtained from IBEX analysis form long narrow tubes in the parameter space. The parameter sets taken from these tubes fit the data almost equally well as the optimum solutions. Therefore we calculated a parameter set that would be acceptable from the view point of IBEX observations and simultaneously would agree with the inflow longitude we derived from the analysis of GAS data. For the inflow longitude 255.3\degr, we calculate the latitude 5.2\degr, speed 25.5 km/s, and temperature 8320 K. Thus, the agreement with the present GAS fitting is reasonable for speed and perhaps latitude, but not for temperature, which comes out even higher than from our GAS data analysis.

\section{Analysis and discussion}
\subsection{Absolute calibration}

\begin{figure*}
 \centering
   \includegraphics[width=18.0cm]{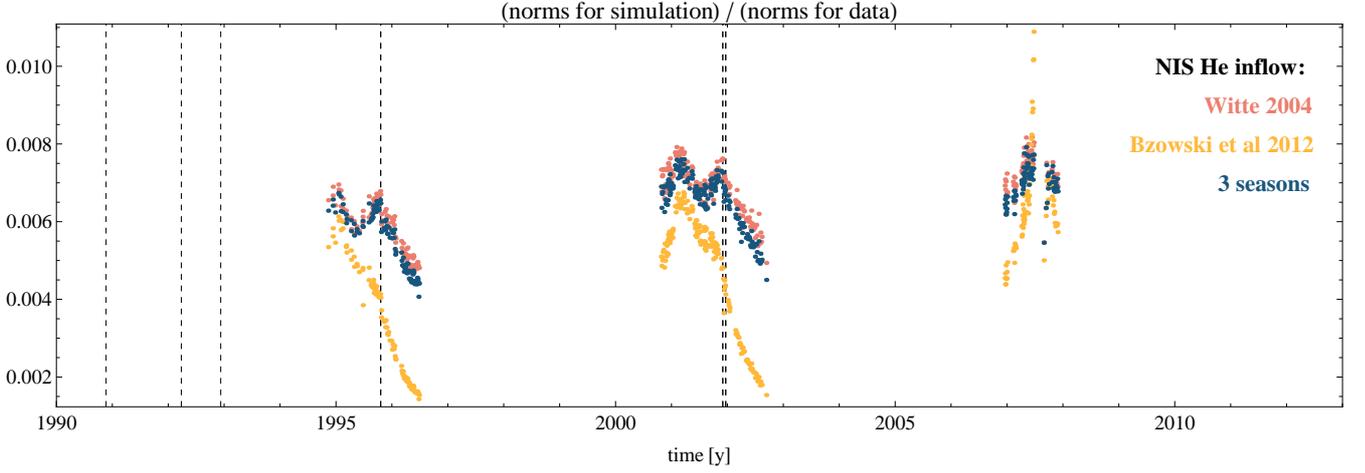}
   \caption{Ratios of normalization factors for simulation results to normalization factors for the data. The normalization factors are by definition adopted as a sum of the signals from all NIS He pixels in the frame (with the background subtracted from the data), for the flow parameters as obtained by \citet{witte:04} from Ulysses (red), by \citet{bzowski_etal:12a} from IBEX (orange), and in the present analysis (blue). The vertical bars mark the times of the detector active surface reprocessing by LiF evaporation.}
    \label{norm.best.fit}
\end{figure*}

The behavior of the normalization factors discussed in the Simulations section is an evidence that the solution we found is self-consistent. The normalization factors for the simulations are proportional to the expected total flux, integrated over the field of view,  modified by the collimator, with the instrument energy response taken into account. The ratios of normalization factors for the simulation to the normalization factors for the data are presented in Fig.~\ref{norm.best.fit}. If the ionization rate adopted is correct, the energy sensitivity does not change with time, and the total density of the gas in front of the heliosphere does not change with time, the season to season averages should be equal. If the model adopted is fully correct, the simulation/data ratio within a season should feature a normally-distributed scatter. In reality, the ratio of normalization factors varies with time by $\sim 10$\% in a systematic way for the original solution from \citet{witte:04}, for the best solution from IBEX, and for the parameter set from the present analysis. While the ratios for the \citet{witte:04} and present solutions closely match each other, the ratio for the parameters from IBEX are different. 

The season-to-season repeatability of the ratio pattern is correlated with the distance of Ulysses from the Sun and thus with the relative velocity between Ulysses and interstellar He gas. We conclude that the relatively small seasonal variation in the ratio is due to imperfections of the energy sensitivity calibration we adopted. The season averages of the ratios show some small systematic change from season to season and we believe that this may be due to an overall small drop in the total sensitivity rather than to a systematically increasing departure of the adopted ionization rate from the true one or a systematic change in the NIS He gas density. 

\subsection{Potential variations of the inflow parameters with time}
\begin{figure*}
 \centering
   \includegraphics[width=18.0cm]{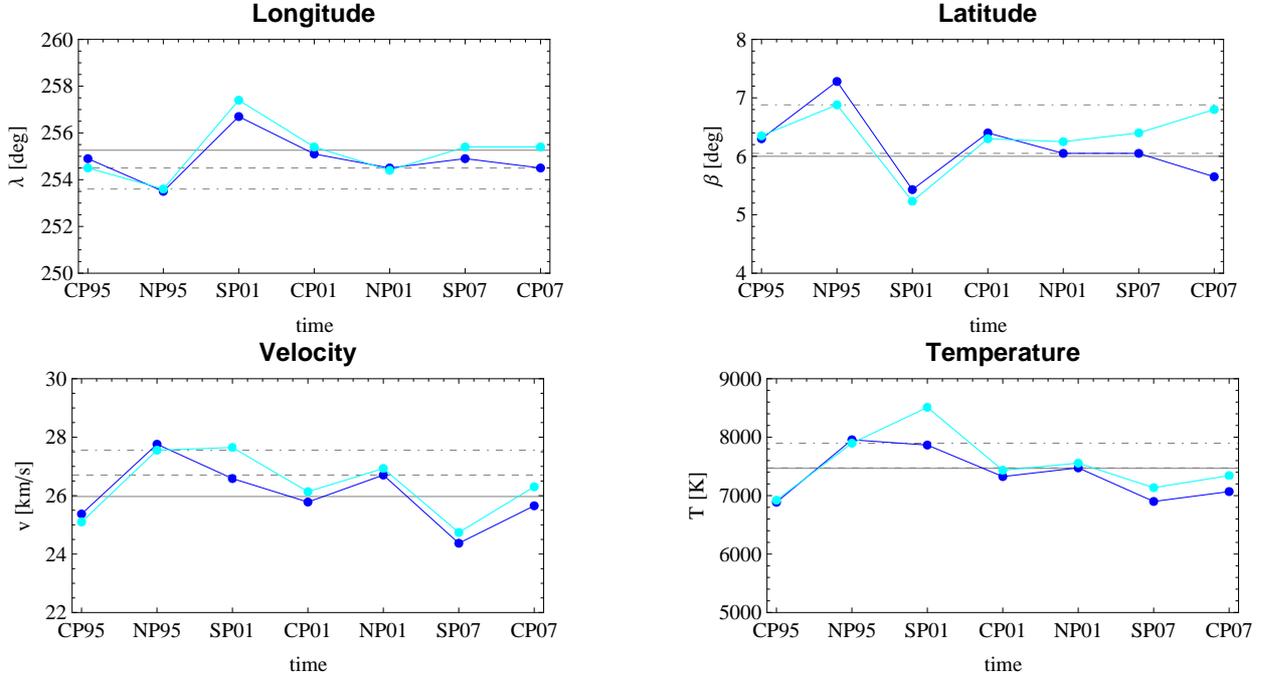}
   \caption{Results of gas inflow parameter fitting for the orbital arcs arranged in a time sequence, shown for both the original and newly-reprocessed data sets. The horizontal axes provide identification of the arcs defined in Fig.~\ref{figUlyCombo}, with the parameters defined in Tables \ref{tabMinChi2Old} (shown as the cyan symbols) and in \ref{tabMinChi2ReProc} (dark-blue). The solid horizontal bars mark the result of optimization for all data from three seasons (see Table \ref{tabMinChi2Oldseasons}), the dashed line the solution with the lowest chi-squared from the reprocessed data set, and the dash-dot line the solutions for the original data set.}
    \label{fig_timeSeriesBoth}
\end{figure*}

The parameters obtained from fitting the three polar orbits separately do not show evidence of any statistically significant change with time (see Table~\ref{tabMinChi2Oldseasons}). Also no change show the parameters obtained from fits to various subsets of the data, discussed further in the text for other purposes. The lack of change with time is illustrated in Fig.~\ref{fig_timeSeriesBoth}, where the fit results are arranged chronologically. The numerical values of the parameters presented in this figure are collected in Tables~\ref{tabMinChi2Old} and \ref{tabMinChi2ReProc}. 

\begin{table}
\caption[]{Inflow longitude, latitude, speed, and temperature of interstellar fitted to the data from individual orbital arcs of Ulysses, defined in Fig.~\ref{figUlyCombo}; data used were in the original form as used by \citet{witte:04}.}
\label{tabMinChi2Old}
\centering
\begin{tabular}{l c c c c}	
\hline \hline
    & $\lambda$[\degr] & $\beta$[\degr] & $v$~[km/s] & $T$[K] \\
\hline
NP95 & 253.6 & 6.88 & 27.555 &  7897  \\
NP01 & 254.4 & 6.25 & 26.930 &  7556  \\
SP01 & 257.4 & 5.23 & 27.647 &  8511  \\
SP07 & 255.4 & 6.40 & 24.739 &  7136  \\
CP95 & 254.5 & 6.35 & 25.097 &  6921  \\
CP01 & 255.4 & 6.30 & 26.136 &  7434  \\
CP07 & 255.4 & 6.80 & 26.302 &  7343  \\
\hline
\end{tabular}

\end{table} 

\begin{table}
\caption[]{Inflow longitude, latitude, speed, and temperature of interstellar He fitted to the data from individual orbital arcs of Ulysses; data reprocessed for the present paper.}
\label{tabMinChi2ReProc}
\centering
\begin{tabular}{l c c c c}	
\hline \hline
    & $\lambda$[\degr] & $\beta$[\degr] & $v$~[km/s] & $T$[K] \\
\hline
NP95 & 253.5 & 7.28 & 27.759 &  7955  \\
NP01 & 254.5 & 6.05 & 26.704 &  7475  \\
SP01 & 256.7 & 5.43 & 27.584 &  7865  \\
SP07 & 254.9 & 6.05 & 24.371 &  6900  \\
CP95 & 254.9 & 6.30 & 25.372 &  6886  \\
CP01 & 255.1 & 6.40 & 26.780 &  7326  \\
CP07 & 254.5 & 5.65 & 26.652 &  7069  \\
\hline
\end{tabular}

\end{table} 

\subsection{Constraining the gas inflow parameters and determining the uncertainty ranges}

It seems that critical for understanding the reason for the discrepancies between the GAS and IBEX results is resolving the question of inflow longitude and temperature. It was argued that the inflow longitude is relatively easy to obtain from measurements of the helium backscatter glow, specifically, from the observations of the cone region. \citet{vallerga_etal:04a} argued they measured the longitude of the inflow in perfect agreement with the GAS result, and their error bars are only 0.5\degr. \citet{bzowski_etal:12a} and \citet{mobius_etal:12a} found that the longitude from GAS is at the border of their acceptable region, but then the gas inflow temperature and speed must be accordingly modified. On the other hand, the temperatures obtained from GAS and IBEX agree very well. But then, if we take the temperature, we cannot change the IBEX-derived longitude and speed so that they agree with the Ulysses result.

\begin{figure}
{\includegraphics[width=9.0cm]{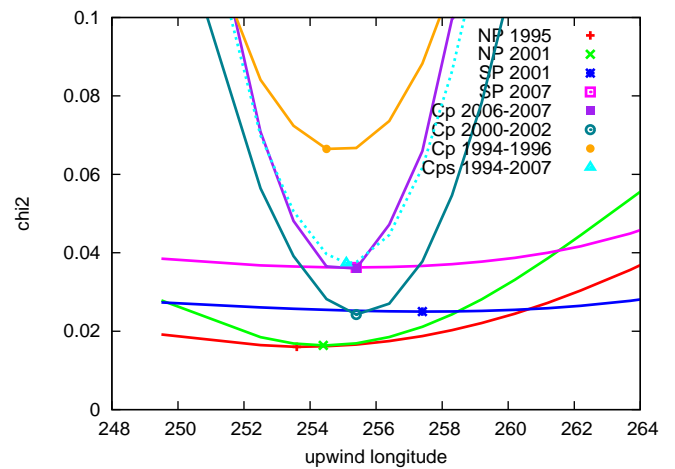}}
\caption{Chi-squared as a function of inflow longitude for fits performed on sections of the data corresponding to the orbital arcs marked in Fig.~\ref{figUlyCombo}. The minima are marked as thick points on the $\chi^2(\lambda)$ lines. The dotted line is the $\chi^2$ fit for the three CP arcs together.}
\label{chi.longg.pole.Cp}
\end{figure}

In attempt to resolve these dilemmas, we checked how well is the inflow longitude constrained by the GAS observations. To that end, we cut the observations from each polar orbit into three exclusive arcs and we fitted the solutions to those arcs independently. The selection of arcs is presented in Fig.~\ref{figUlyCombo}. We made the data selection carefully so that they are taken on identical intervals of heliolatitude, and fitted the parameters orbit by orbit. Common between the three polar orbits is the CP (common part) arc. In addition, there are north (NP) and south (SP) polar arcs. We performed fits both to the original data and to the data reprocessed for the needs of this paper. The distribution of data frames within the data subsets is shown in Fig,~\ref{figFrames}. The chi-squared minimization lines are shown in Fig.~\ref{chi.longg.pole.Cp} and the numerical values for the optimum parameters are presented in Fig.~\ref{fig_timeSeriesBoth} as well as in Tables~\ref{tabMinChi2Old} and \ref{tabMinChi2ReProc}, for the original and reprocessed data, respectively.

\begin{figure*}
\begin{tabular}{lll}
\includegraphics[width=.3\textwidth]{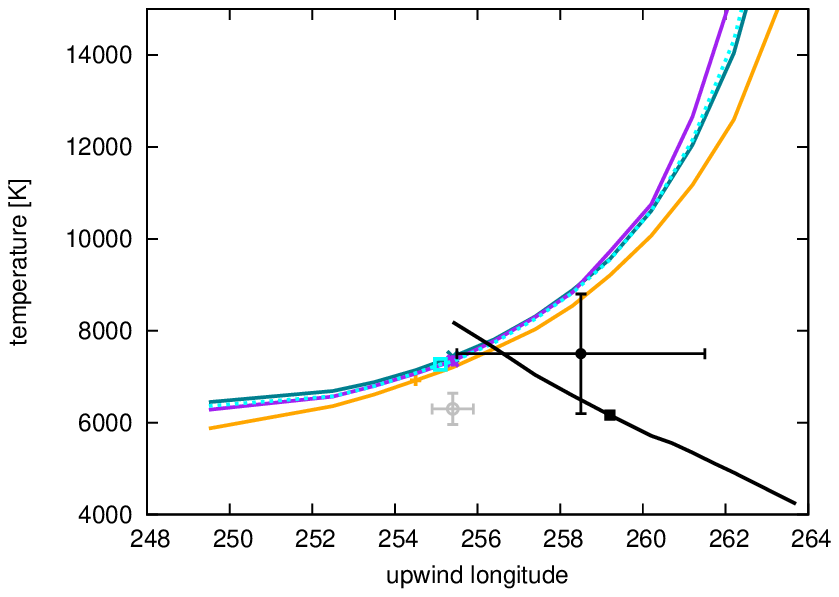} & 
\includegraphics[width=.3\textwidth]{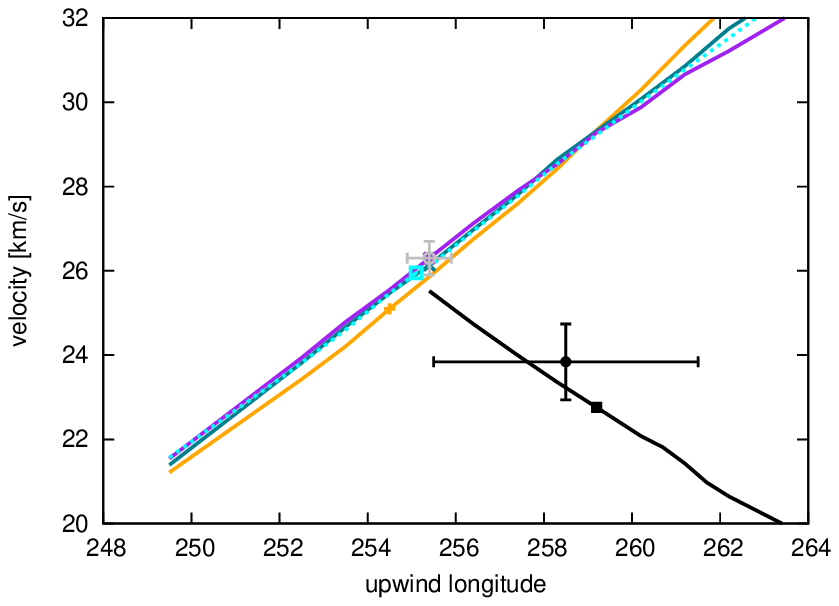} & 
\includegraphics[width=.3\textwidth]{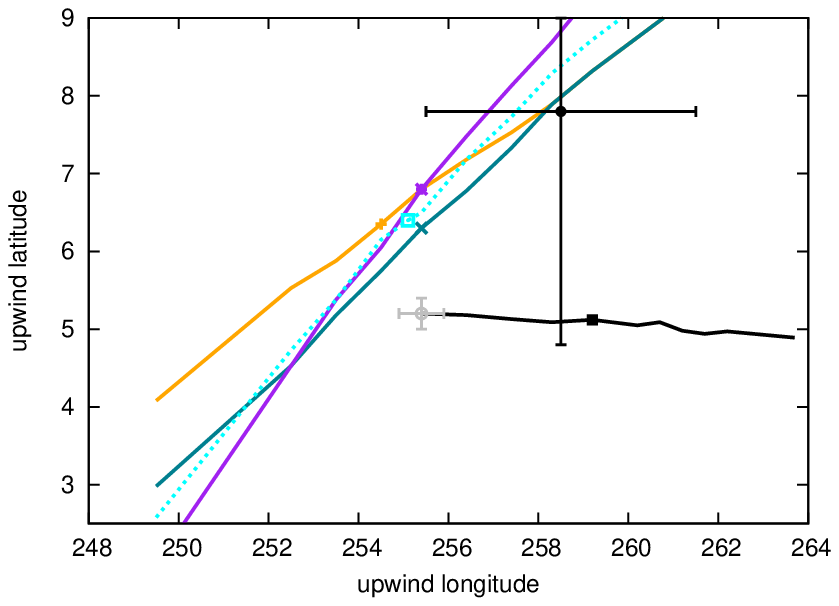}  \\
\includegraphics[width=.3\textwidth]{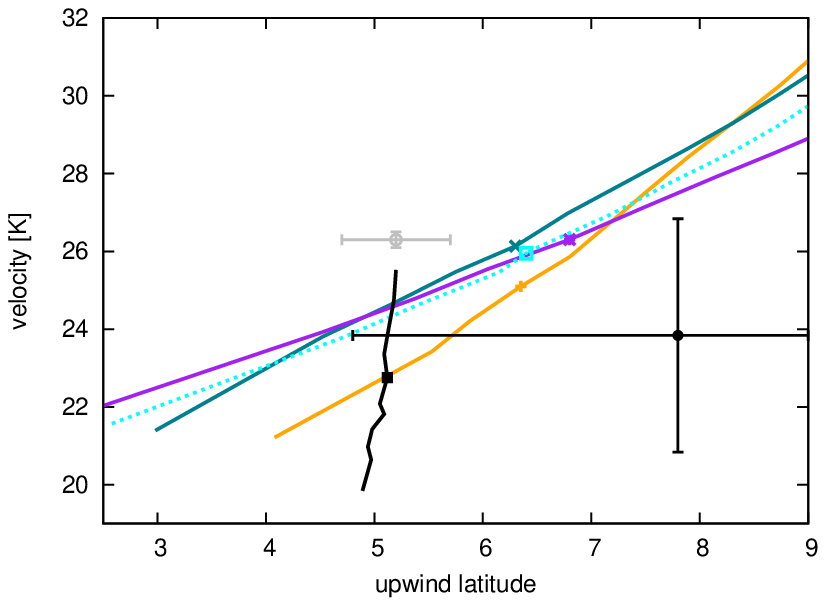} & 
\includegraphics[width=.3\textwidth]{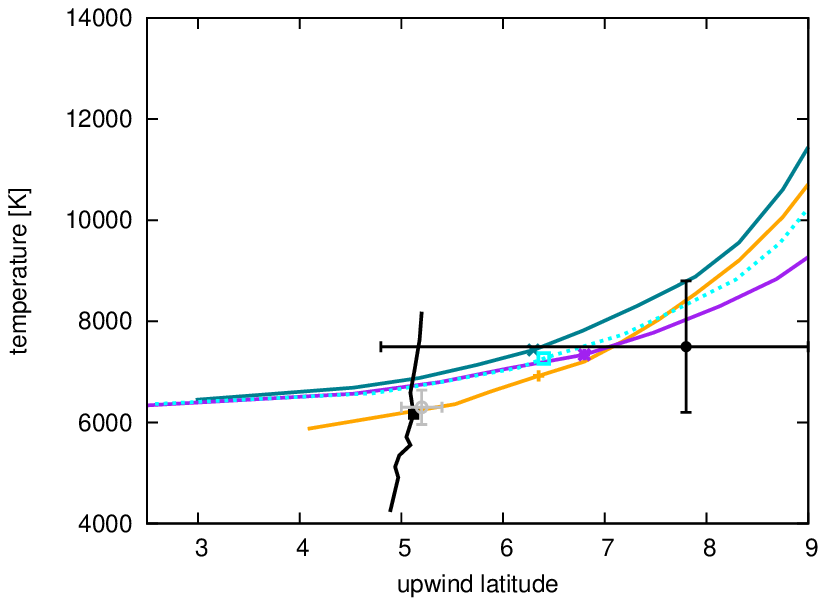} & 
\includegraphics[width=.3\textwidth]{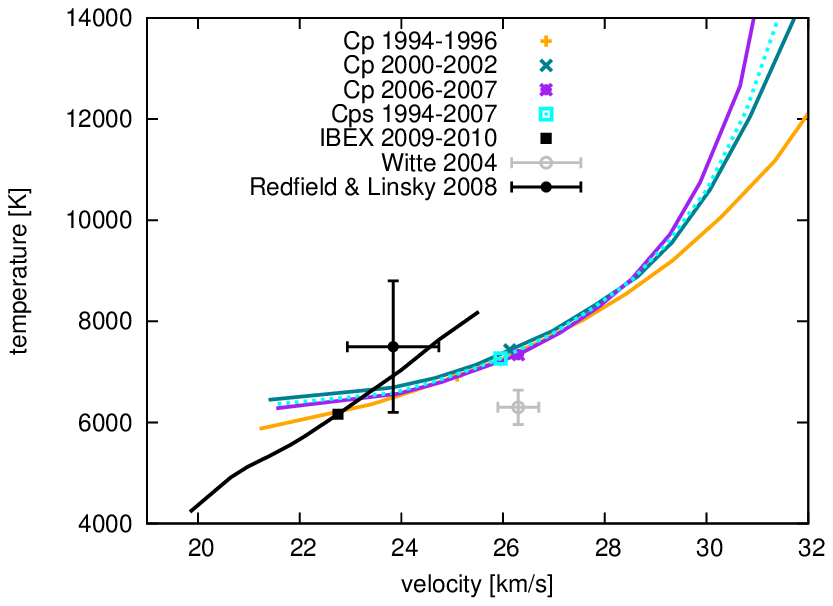}  \\
\end{tabular}
\caption{Parameter correlation lines, shown pairwise for the three common-part (CP) scans (color lines). The corresponding $\chi^2(\lambda)$ lines for these scans are shown in Fig.~\ref{chi.longg.pole.Cp}. The small gray cross is the solution found by \citet{witte:04}, with error bars. The black cross marks the Local Interstellar Cloud flow parameters found from interstellar absorption lines by \citet{redfield_linsky:08a}, the black line is the acceptable parameter line obtained from IBEX by \citet{bzowski_etal:12b}, with the dark spot marking the best solution. The color dots mark the optimum sets found for individual lines.}
\label{par.eclip}
\end{figure*} 

As already noted, the results are in agreement between the seasons, i.e., they do not show any clear, statistically significant temporal trends. The differences between the results obtained from the original and reprocessed data sets are minimal; more discussion on the possible reasons for those small differences is provided further on in the paper.

\begin{figure*}
\begin{tabular}{lll}
\includegraphics[width=.3\textwidth]{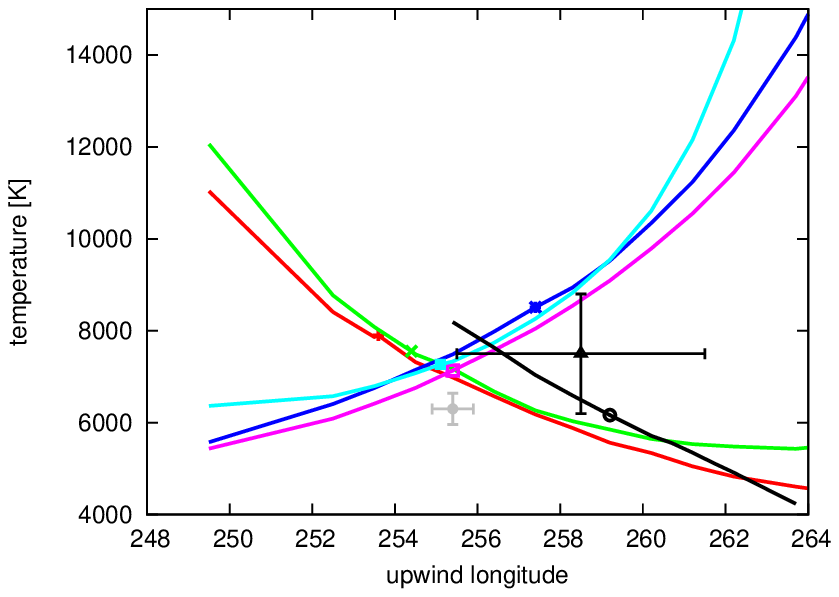} & 
\includegraphics[width=.3\textwidth]{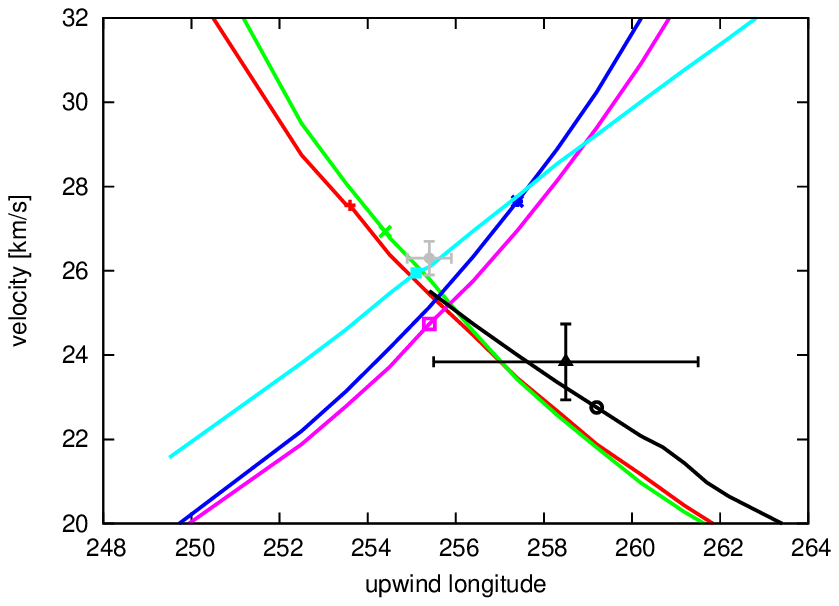} & 
\includegraphics[width=.3\textwidth]{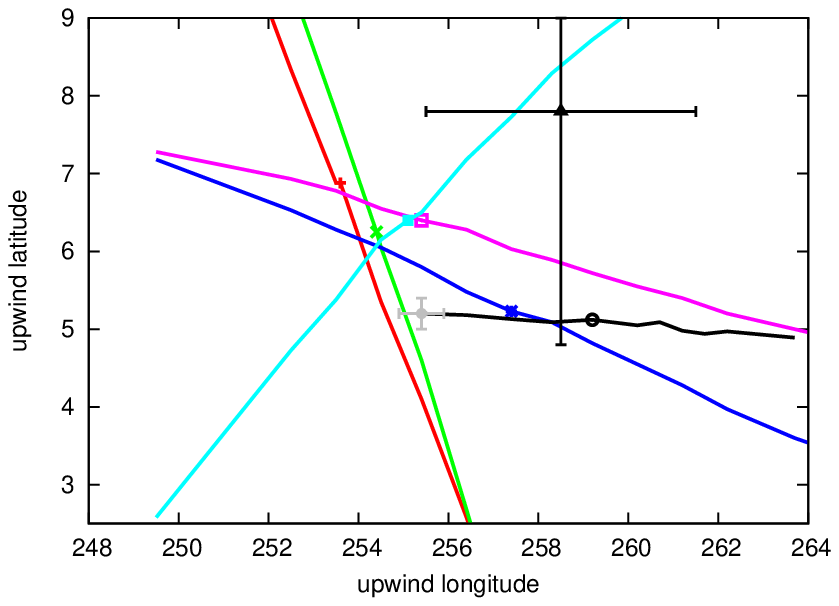}  \\
\includegraphics[width=.3\textwidth]{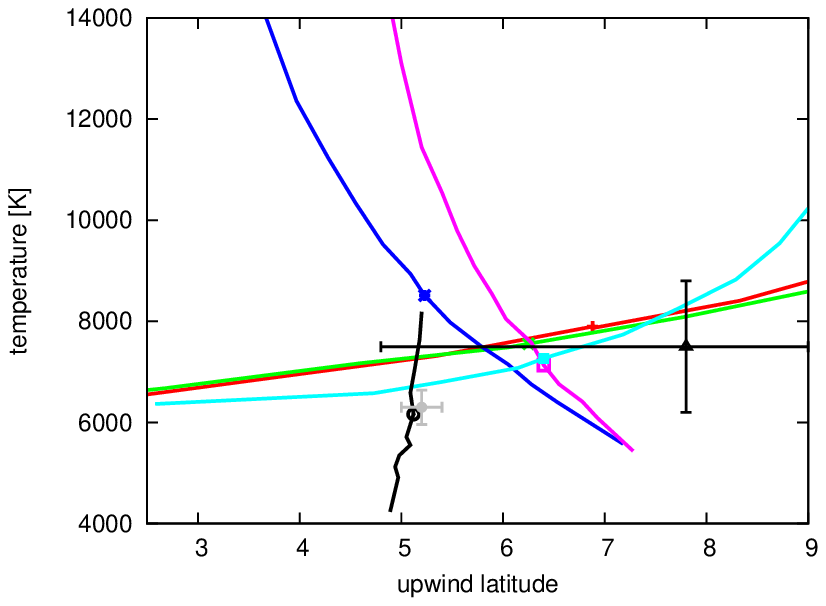} & 
\includegraphics[width=.3\textwidth]{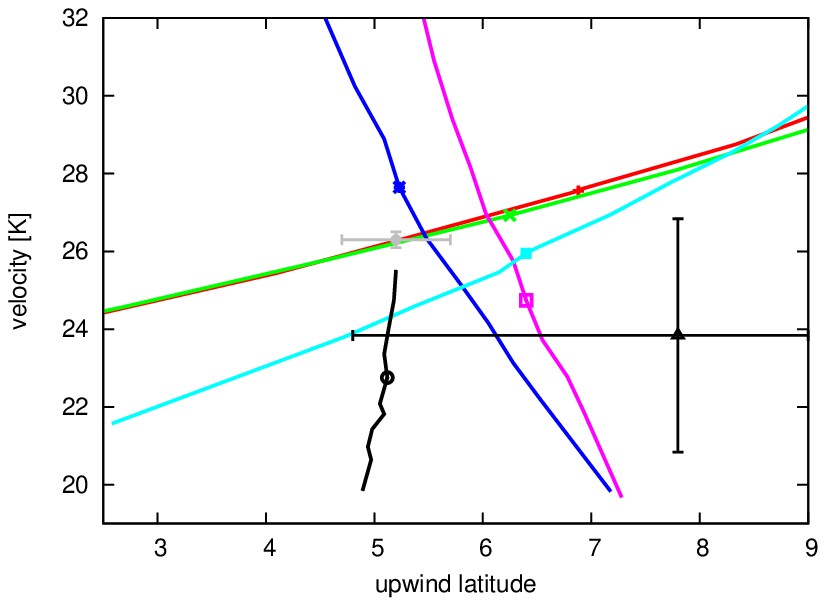} & 
\includegraphics[width=.3\textwidth]{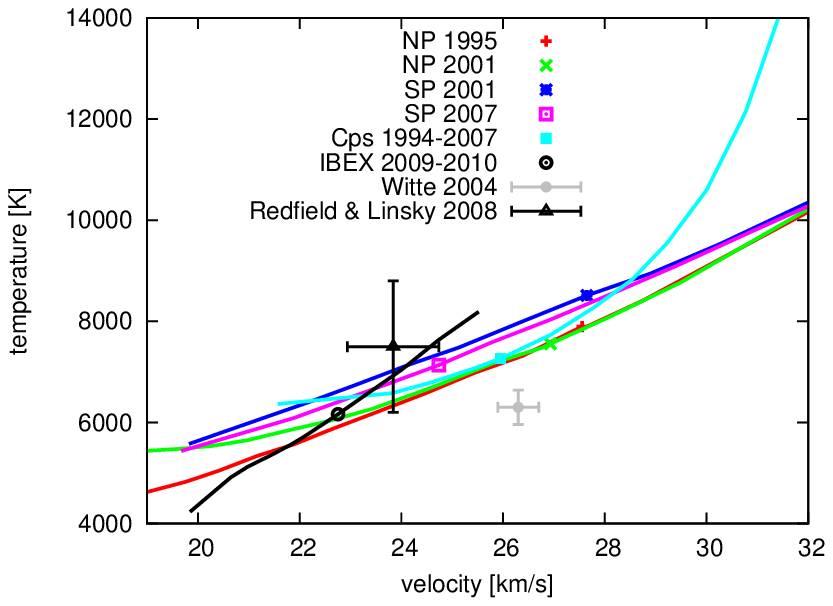}  \\
\end{tabular}
\caption{Parameter correlation lines, shown pairwise for the polar data subsets illustrated in Fig. \ref{figUlyCombo}, with $\chi^2(\lambda)$ shown in Fig.~\ref{chi.longg.pole.Cp}. The small gray cross with error bars is the solution found by \citet{witte:04}. The large black cross marks the LIC flow parameters found from analysis of interstellar absorption lines by \citet{redfield_linsky:08a}, the black line is the acceptable parameter line obtained from IBEX by \citet{bzowski_etal:12b}, with the dark spot marking the best solution. In addition to the correlation lines from the polar arcs (north NP and south SP), parameter correlation lines from the fit to the three CP scans together are shown. The color dots mark the optimum parameter sets found for individual arcs.}
\label{par.pole}
\end{figure*} 

The inflow parameters can hopefully be better constrained by the common part scans. Adopting somewhat arbitrarily as limiting value the minimum plus 0.02 for a given data subset we obtain that the acceptable longitudes of the inflow direction vary from $\sim 253\degr$ to 258\degr (see Fig.~\ref{chi.longg}), i.e., the boundary is not far from the IBEX values. 

\begin{figure*}
\begin{tabular}{rrr}
\includegraphics[width=.3\textwidth]{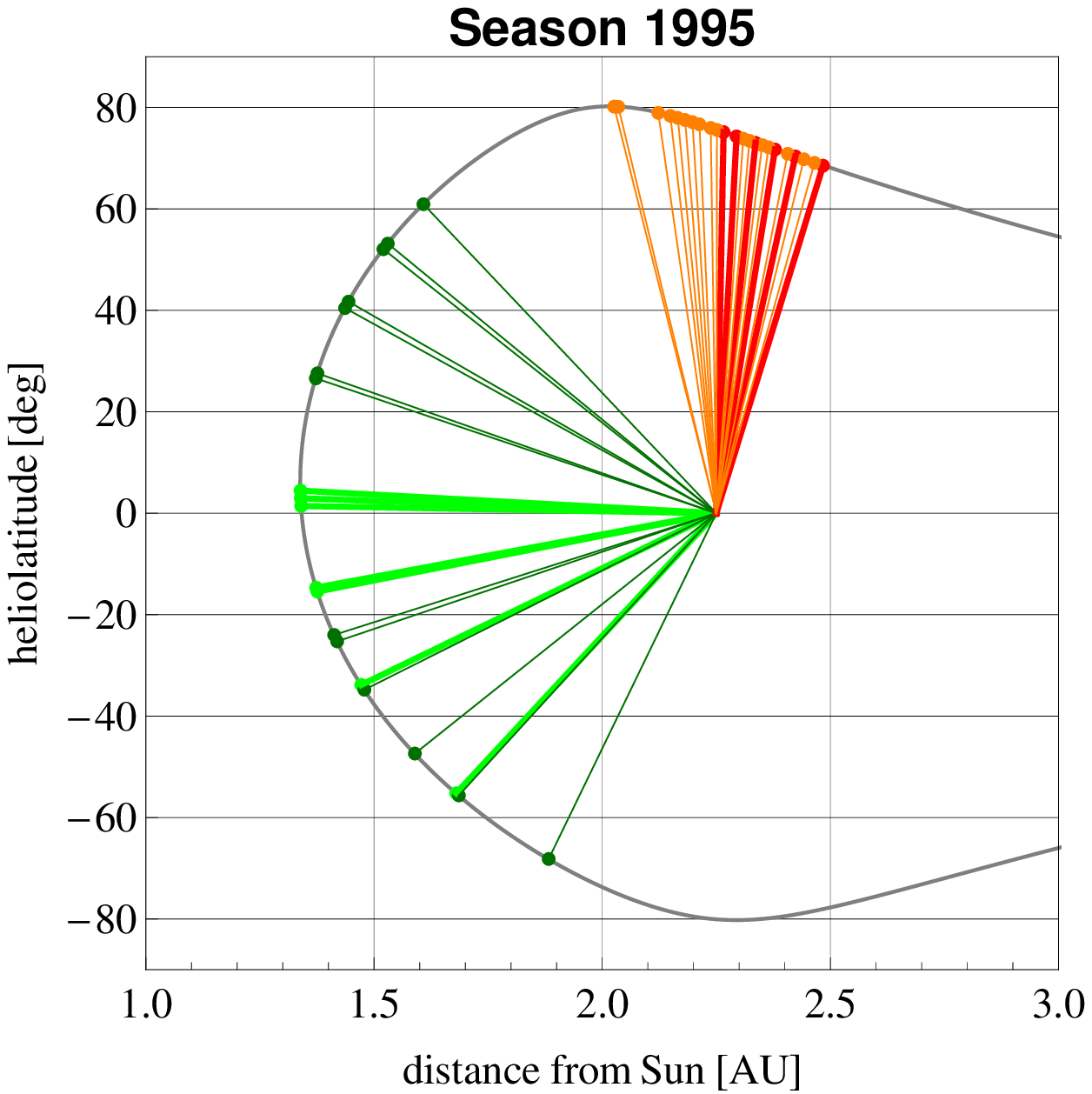} & 
\includegraphics[width=.3\textwidth]{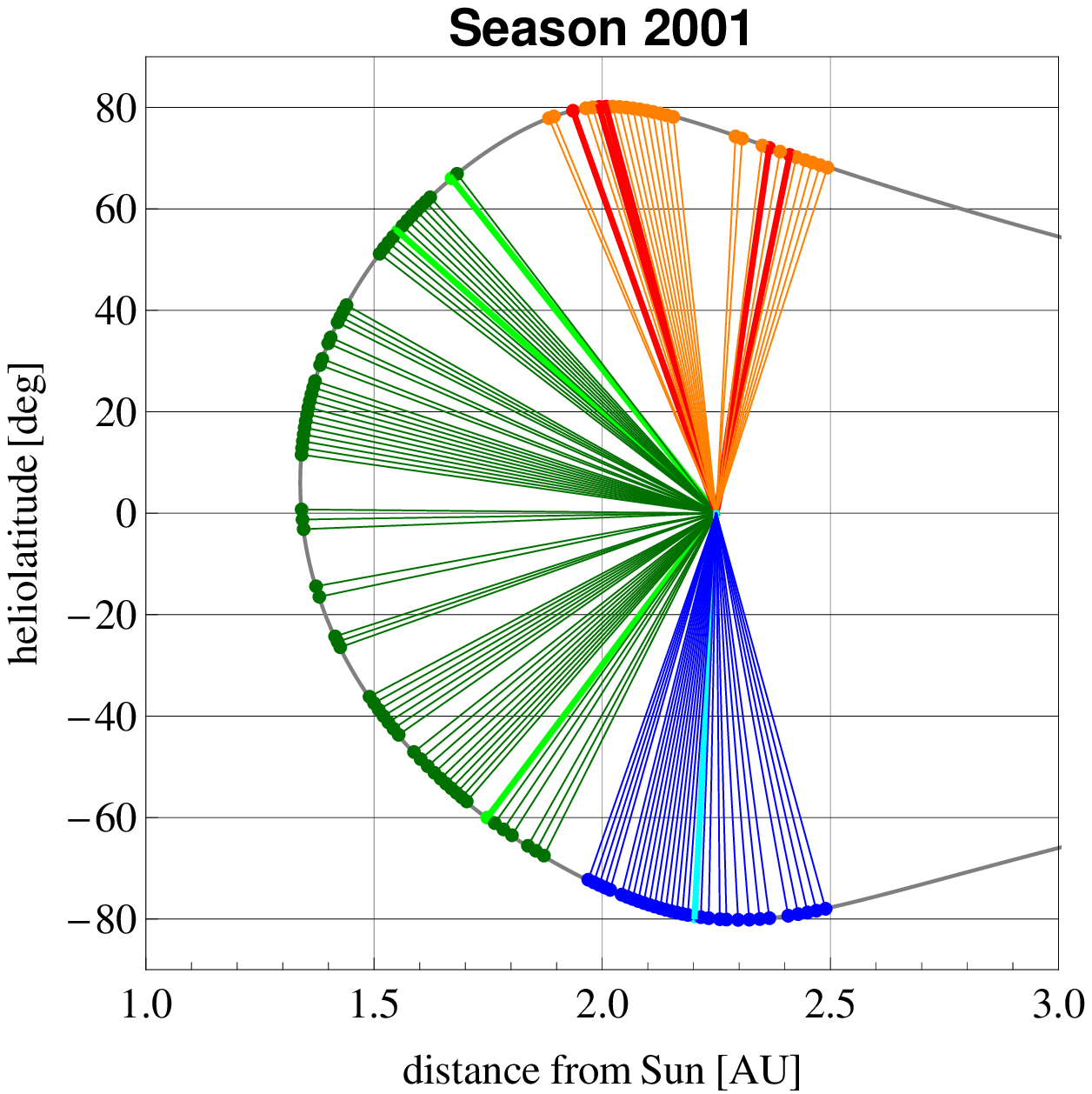} & 
\includegraphics[width=.3\textwidth]{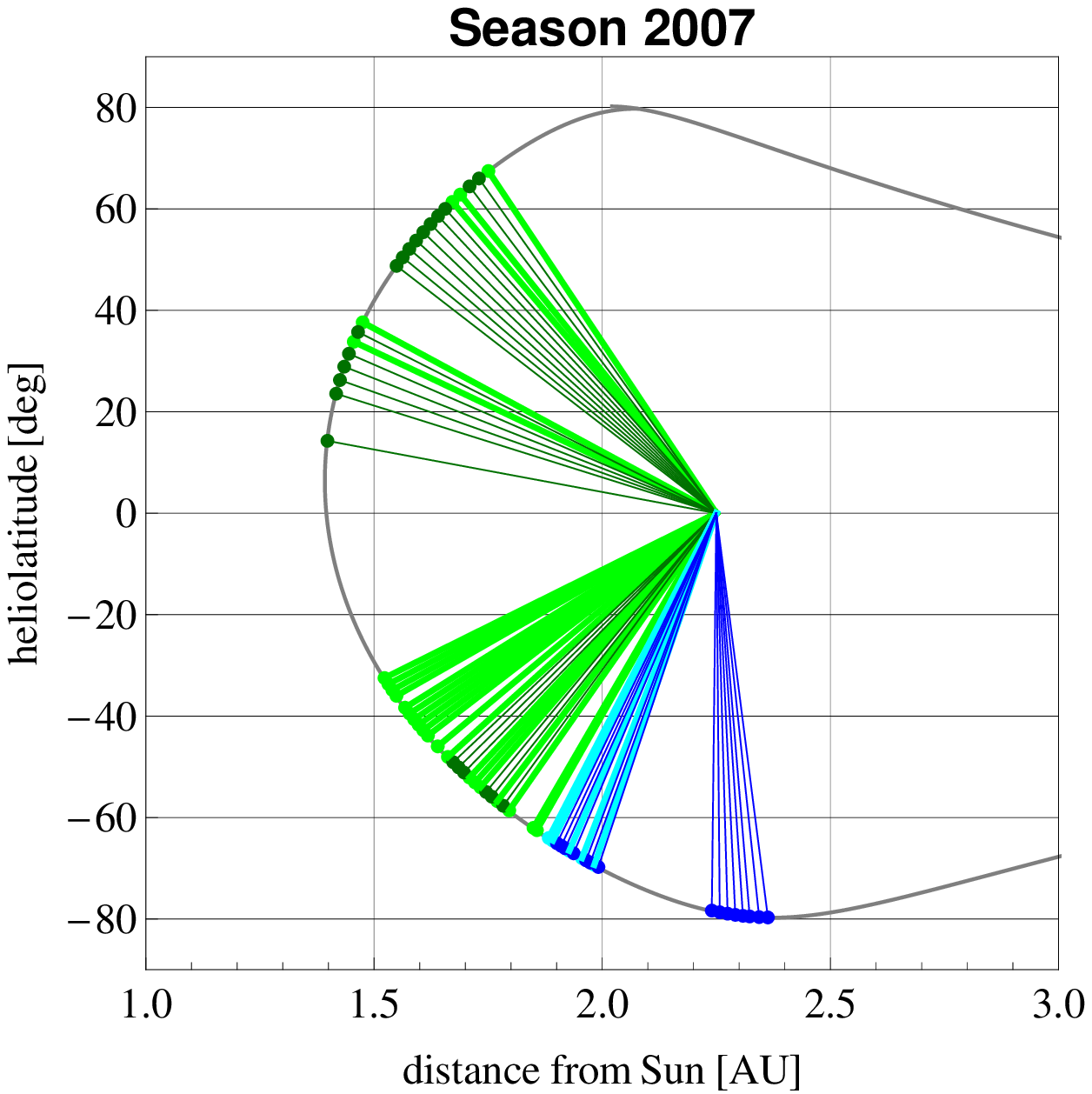}  \\
\end{tabular}
\caption{Distributions of the data frames taken to the analysis on the three Ulysses polar orbits. The brighter colors mark the data frames that were rejected in the reprocessed data set.}
\label{figFrames}
\end{figure*} 

\begin{figure*}
\begin{tabular}{lll}
\includegraphics[width=.3\textwidth]{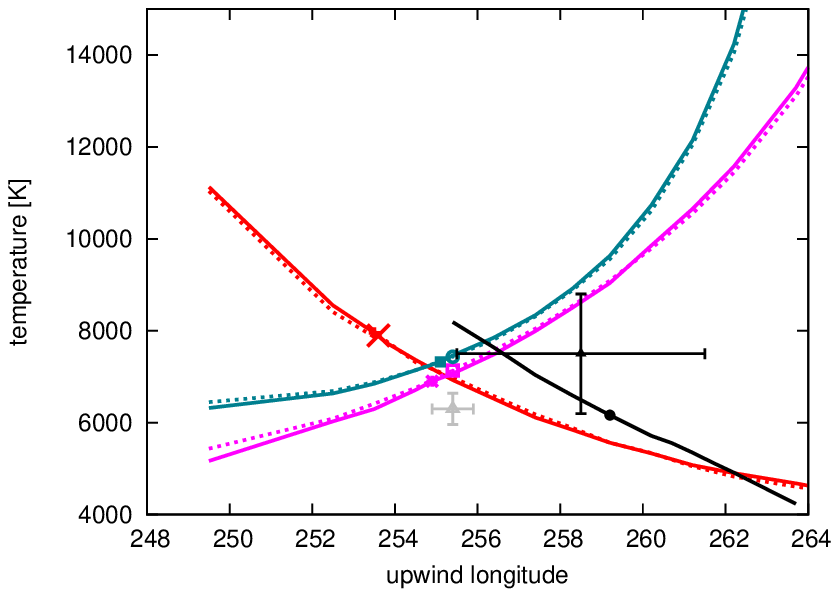} & 
\includegraphics[width=.3\textwidth]{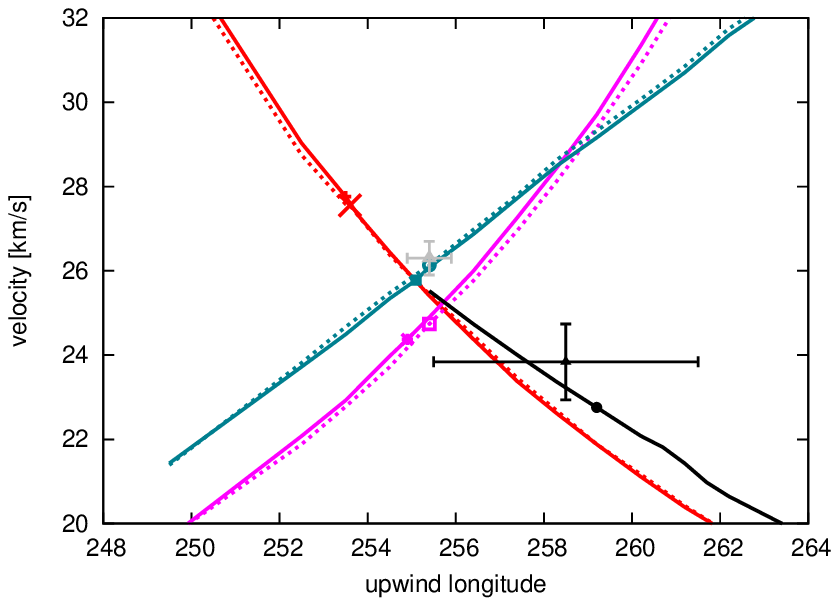} & 
\includegraphics[width=.295\textwidth]{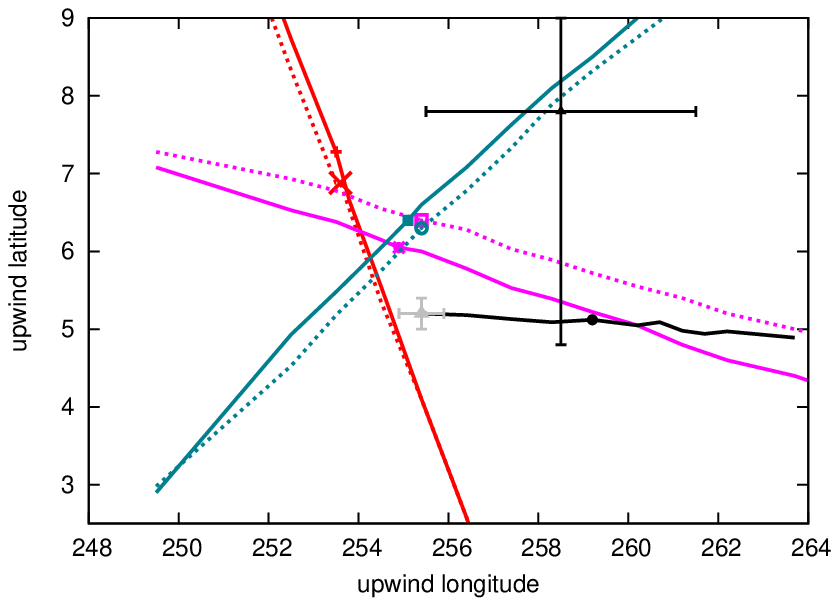}  \\
\includegraphics[width=.3\textwidth]{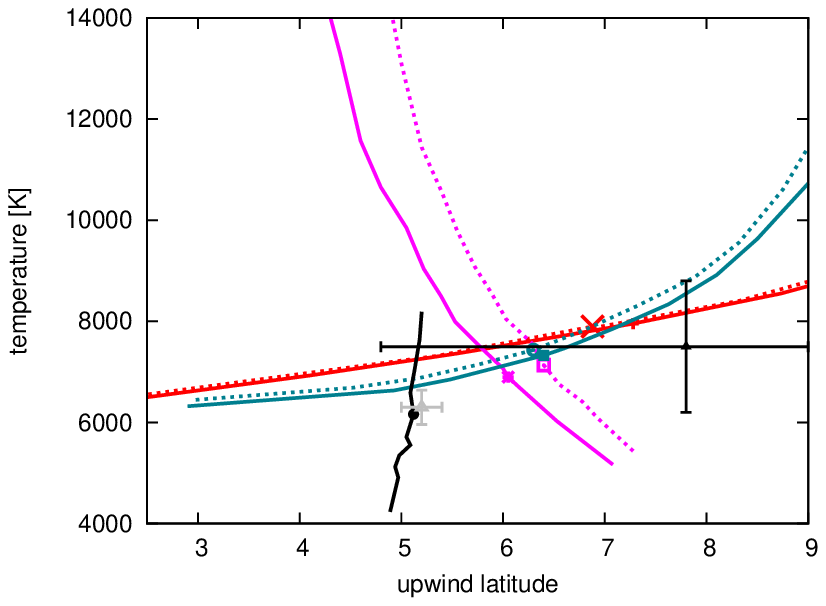} & 
\includegraphics[width=.3\textwidth]{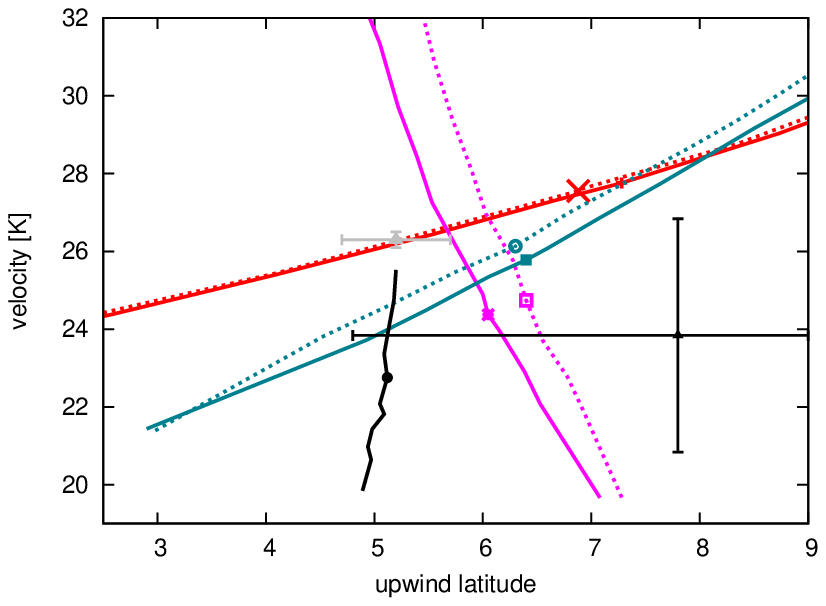} & 
\includegraphics[width=.31\textwidth]{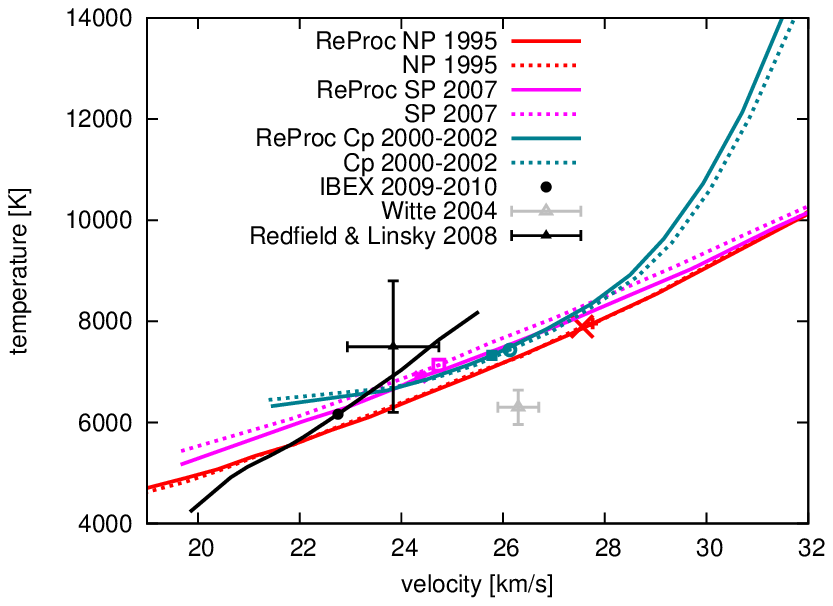}  \\
\end{tabular}
\caption{Parameter correlation lines obtained from analysis of the original and reprocessed data set, shown for the north and south polar passes, and for the common parts of the orbits, for all three Ulysses polar orbits. The data selection is presented in Fig.~\ref{figFrames}. Solid lines represent lines from the reprocessed data, broken lines from the original data. The black line represents the acceptable parameter line from IBEX \citep{bzowski_etal:12a}, the black cross the interstellar gas parameters obtained by \citet{redfield_linsky:08a} from analysis of interstellar absorption lines (with uncertainties), the gray cross marks the NIS He parameters found by \citet{witte:04}, converted to J2000. The dots with color matching the line colors mark the parameters for which $\chi^2(\lambda)$ minimum was found for a given subset of data.}
\label{par.Comp}
\end{figure*} 

Similarly as it was demonstrated for the IBEX observation conditions, also for Ulysses the parameter values that belong to the $\chi^2(\lambda)$ lines in Fig.~\ref{chi.longg.pole.Cp} are related to each other, as illustrated in Fig.\ref{par.eclip}. The lines from equivalent arcs are very close to each other from orbit to orbit. They are in marginal agreement with the acceptable parameter regions obtained from IBEX except for the longitude -- latitude parameter pair. They are in agreement within the error bars with the LIC flow parameters obtained by \citet{redfield_linsky:08a} from analysis of interstellar absorption lines, in this case except for the longitude -- velocity pair. Interestingly, they are outside the error bars reported by \citet{witte:04} for the temperature -- speed pair, which may suggest the uncertainty of this result was assessed too optimistically.  

Another selection for the analysis were the polar arcs. This analysis shows that the data from the relatively short polar arcs only weakly constrain the parameters when analyzed individually: the minima of $\chi^2(\lambda)$ are relatively shallow, especially for the south-pole arcs, where the background was higher and less homogeneous than for the north polar passes (cf Fig.~\ref{chi.longg.pole.Cp}). The parameter correlation lines for this case are shown in Fig.~\ref{par.pole}. While the lines from the south polar arc are quite similar to the common-part latitude scan lines, the lines from the north polar arc are very different, intersecting with the former ones in relatively well constrained regions. Regrettably, the complement of the north and south polar arcs is available only for one orbit, namely the second one. The parameter lines from the two north passes (from the first and second Ulysses orbits) agree with each other very well. In the case of the south arcs (from the second and third orbit), the agreement is less tight. 

Basically, the intersections can be used to better constrain the parameters. However, close inspection of Fig.~\ref{par.pole} shows that constraints from different parameter pairs are somewhat different. Moreover, it seems that the locations of the polar parameter lines in the parameter space depend on the data selection relatively strongly. We draw this conclusion from comparison of the results obtained from analysis of the original and reprocessed data. 

The number of data frames in the reprocessed data set is lower than in the original, especially in the south pole subset. A comprehensive view of the locations on the Ulysses orbit from which individual data frames were taken is provided in Fig.~\ref{figFrames}. Analysis of this figure suggests that the most of the rejected data frames are in the south limb of the arc of the 2007 CP scan and in the 2007 south polar pass. Comparison of the parameter correlation lines obtained from the reprocessed data set, presented in Fig.~\ref{par.Comp}, with the parameter lines from the original data set tells us that the parameter lines from the arcs where almost all data frames were retained in the reprocessed data set are almost identical with the original ones. This is evident from the north polar passes. 

\begin{figure*}
\begin{tabular}{lll}
\includegraphics[width=0.3\textwidth]{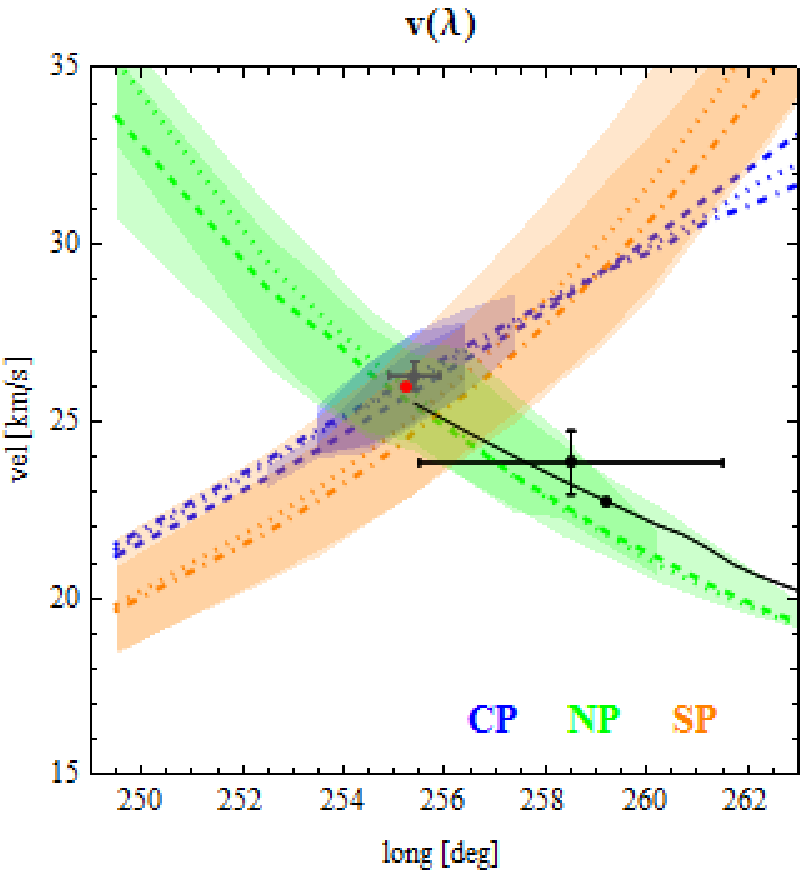} &
\includegraphics[width=0.32\textwidth]{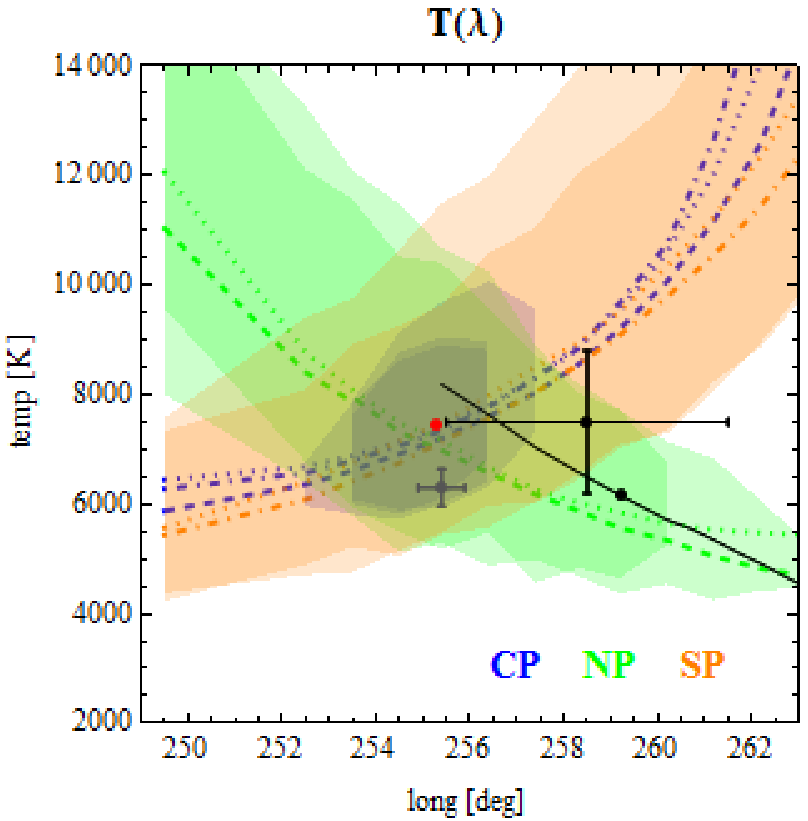} &
\includegraphics[width=0.3\textwidth]{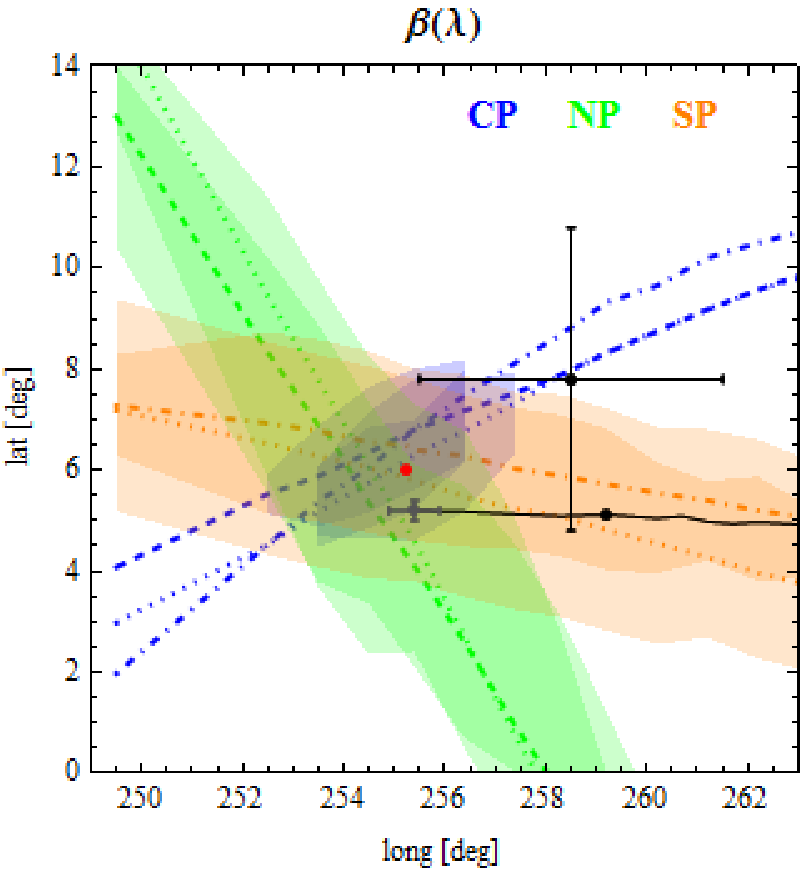} \\
\end{tabular}
\caption{Acceptable parameter regions based on various cuts of the data. The criterion is that $\chi^2(\lambda)$ is within 0.02 from the minimum for the respective cut and season. The colored regions correspond to the orbital arcs for individual seasons. The green region corresponds to the north polar passes (NP), the orange region to the south polar passes (SP), the blue region to the common-part scans (CP). A higher intensity of a color means an intersection of acceptable regions obtained from different seasons for a given orbital arc. The color lines of different styles mark the parameter correlation lines from Fig.~\ref{par.pole} for different seasons. The black line is the acceptable parameter line from IBEX \citep{bzowski_etal:12a}, with the best solution marked with the black point. The red dot marks the parameters obtained from $\chi^2$ minimization of all GAS/Ulysses data available (this work). The big black cross marks the LIC flow parameters from \citet{redfield_linsky:08a}, the small gray cross the NIS He inflow parameters from \citet{witte:04}.}
\label{figContours}
\end{figure*}

Thus, reprocessing the north-pole arc data had very little effect on the results, i.e., the original data set provided accurate results. However, the lines from the south polar passes differ for some parameter pair combinations and we believe that this is because of data selection. The differences between the original and reprocessed data set are visible especially well for the latitude-speed, latitude-temperature, and, importantly, longitude-latitude pairs. The rejected data frames cluster in one region of the south polar arc and this results in a shift of the acceptable parameter region in the parameter space.

Thus we conclude that the data selection has a significant influence on the conclusions drawn from the analysis of data subsets. While the results are self-consistent, we cannot concede that analysis of the polar arcs provides us with tighter constraints on the acceptable parameters than the analysis of chi-squared, the quality of the two types of constraints is comparable. 

Based on this insight, we adopt as the best fitting the parameters provided in the Results section, obtained from the analysis of the full data set. For the uncertainty range, we take the intersection of the acceptable parameter regions from the polar and CP scans, illustrated in Fig.~\ref{figContours}. For longitude and speed, we obtain $254.2\degr < 255.3\degr < 256.5\degr$ and $24.5 < 26.0 < 27$~km/s, respectively. With these constraints, the temperature is in the range $5500 < 7500 < 9000$~K, this parameter being constrained the weakest. The latitude is in the range $5.0\degr < 6.0\degr < 7.0\degr$. These constraints are of course approximate since, as evident from Fig.~\ref{figContours}, the acceptable region, marked as the darkest shaded region in the plots, is comparably complex in shape. The original solution obtained by \citet{witte:04} is inside the uncertainties of the present one, even though it seems located rather close to the boundaries. On the other hand, while the acceptable parameter regions obtained from this study and from IBEX seem to partly overlap, the optimum solution found now and the solution found from IBEX analysis still differ. The LIC flow vector obtained by \citet{redfield_linsky:08a} is outside the acceptable parameter region obtained now.

\section{Summary and conclusions}

We have performed an extensive reanalysis of observations from the GAS experiment onboard Ulysses. We analyzed again the original data set, previously analyzed by \citet{witte:04}, and extended the analysis to the last Ulysses orbit in 2007, previously not analyzed. In addition, we reprocessed the original data set, trying to improve the pointing of the instrument boresight and to eliminate the portions of the data with high background. The results of the analysis are in good agreement with the previous ones, except for the temperature. The former result is within the uncertainty range of the present one, even though the uncertainty range as we see now is much larger than previously. The most likely values for the interstellar He inflow parameters are longitude 255.3\degr, latitude 6.0\degr, speed 26.0~km/s, and temperature 7500~K, the latter one being significantly higher than the originally-derived $\sim 6300$~K. We do not see any evidence for a statistically significant temporal change in the inflow parameters for the duration of the Ulysses mission. We do observe a small systematic change in the observed intensity, which we interpret as most likely due to a small gradual reduction in the instrument sensitivity. We also see a residual variation in the simulation/data ratio of the normalization factors, which is repeatable season to season and correlated with the solar distance. We believe this small residual discrepancy may be due to small imperfections in the energy calibration used in our study. We have verified, however, that adopting alternative calibration discussed in original papers \citep[e.g.,][]{banaszkiewicz_etal:96a} does not change the results significantly.
The uncertainty ranges of the present GAS result and the results of IBEX-Lo data analysis by \citet{bzowski_etal:12a} and \citet{mobius_etal:12a} marginally overlap, but the most likely values remain discrepant. Within the uncertainties, the IBEX results are in agreement with the determination of LIC flow parameters based on interstellar absorption lines by \citet{redfield_linsky:08a}, which marginally overlap with the uncertainties of the GAS-derived parameters. The discrepancies between these three determinations are probably the best currently available estimate of the uncertainty in the flow vector and temperature of neutral gas in the interstellar medium surrounding the Sun.

\begin{acknowledgements}
The authors gratefully acknowledge support from Polish National Science Center grant 2012-06-M-ST9-00455. This research was partly carried out within the program of International Space Science Institute Working Team 223 {\em Spatial and temporal studies of the heliospheric interaction with the local interstellar medium from SOHO/SWAN UV, IBEX neutral atom, and ACE and STEREO pickup ion observations}.
\end{acknowledgements}

\bibliographystyle{aa}
\bibliography{iplbib}

\end{document}